\documentstyle[12pt]{article}
\def\newpic#1{}
\def\d{\partial}

\def\bea{\begin{eqnarray}}
\def\eea{\end{eqnarray}}

\def\beq{\begin{equation}}
\def\eeq{\end{equation}}
\def\ba{\beq\new\begin{array}{c}}
\def\ea{\end{array}\eeq}
\def\be{\ba}
\def\ee{\ea}

\def\f{1\over}

\makeatletter
\newdimen\normalarrayskip 
\newdimen\minarrayskip 
\normalarrayskip\baselineskip
\minarrayskip\jot
\newif\ifold \oldtrue \def\new{\oldfalse}
\def\arraymode{\ifold\relax\else\displaystyle\fi} 
\def\eqnumphantom{\phantom{(\theequation)}} 
\def\@arrayskip{\ifold\baselineskip\z@\lineskip\z@
\else
\baselineskip\minarrayskip\lineskip2\minarrayskip\fi}
\def\@arrayclassz{\ifcase \@lastchclass \@acolampacol \or
\@ampacol \or \or \or \@addamp \or
\@acolampacol \or \@firstampfalse \@acol \fi
\edef\@preamble{\@preamble
\ifcase \@chnum
\hfil$\relax\arraymode\@sharp$\hfil
\or $\relax\arraymode\@sharp$\hfil
\or \hfil$\relax\arraymode\@sharp$\fi}}
\def\@array[#1]#2{\setbox\@arstrutbox=\hbox{\vrule
height\arraystretch \ht\strutbox
depth\arraystretch \dp\strutbox
width\z@}\@mkpream{#2}\edef\@preamble{\halign
\noexpand\@halignto
\bgroup \tabskip\z@ \@arstrut \@preamble \tabskip\z@ \cr}%
\let\@startpbox\@@startpbox \let\@endpbox\@@endpbox
\if #1t\vtop \else \if#1b\vbox \else \vcenter \fi\fi
\bgroup \let\par\relax
\let\@sharp##\let\protect\relax
\@arrayskip\@preamble}
\def\eqnarray{\stepcounter{equation}%
\let\@currentlabel=\theequation
\global\@eqnswtrue
\global\@eqcnt\z@
\tabskip\@centering
\let\\=\@eqncr
$$%
\halign to \displaywidth\bgroup
\eqnumphantom\@eqnsel\hskip\@centering
$\displaystyle \tabskip\z@ {##}$%
\global\@eqcnt\@ne \hskip 2\arraycolsep
$\displaystyle\arraymode{##}$\hfil
\global\@eqcnt\tw@ \hskip 2\arraycolsep
$\displaystyle\tabskip\z@{##}$\hfil
\tabskip\@centering
&{##}\tabskip\z@\cr}
\begingroup\ifx\undefined\newsymbol \else\def\input#1 {\endgroup}\fi
\newfont{\hr}{msbm10}
\newfont{\ams}{msam10}

\font\numbers=cmss12
\font\upright=cmu10 scaled\magstep1
\def\stroke{\vrule height8pt width0.4pt depth-0.1pt}
\def\topfleck{\vrule height8pt width0.5pt depth-5.9pt}
\def\botfleck{\vrule height2pt width0.5pt depth0.1pt}
\def\Zmath{\vcenter{\hbox{\numbers\rlap{\rlap{Z}\kern 0.8pt\topfleck}\kern
2.2pt \rlap Z\kern 6pt\botfleck\kern 1pt}}}
\def\Qmath{\vcenter{\hbox{\upright\rlap{\rlap{Q}\kern
3.8pt\stroke}\phantom{Q}}}}
\def\Nmath{\vcenter{\hbox{\upright\rlap{I}\kern 1.7pt N}}}
\def\Cmath{\vcenter{\hbox{\upright\rlap{\rlap{C}\kern
3.8pt\stroke}\phantom{C}}}}
\def\Rmath{\vcenter{\hbox{\upright\rlap{I}\kern 1.7pt R}}}
\def\Z{\ifmmode\Zmath\else$\Zmath$\fi}
\def\Q{\ifmmode\Qmath\else$\Qmath$\fi}
\def\N{\ifmmode\Nmath\else$\Nmath$\fi}
\def\C{\ifmmode\Cmath\else$\Cmath$\fi}
\def\R{\ifmmode\Rmath\else$\Rmath$\fi}

\newcounter{app}

\def\app{\setcounter{equation}{0}
\def\theequation{\Alph{app}.\arabic{equation}}\par
\addvspace{4ex}
\@afterindentfalse
\secdef\@app\@dapp}

\newcommand\@app{\@startsection {app}{1}{0ex}%
{-3.5ex \@plus -1ex \@minus -.2ex}%
{2.3ex \@plus.2ex}%
{\normalfont\Large\bf}}
\def\@dapp#1{%
{\parindent \z@ \raggedright \bf #1}\par\nobreak}
\def\l@app#1#2{\ifnum \c@tocdepth >\z@
\addpenalty\@secpenalty
\addvspace{1.0em \@plus\p@}%
\setlength\@tempdima{8em}%
\begingroup
\parindent \z@ \rightskip \@pnumwidth
\parfillskip -\@pnumwidth
\leavevmode \bfseries
\advance\leftskip\@tempdima
\hskip -\leftskip
#1\nobreak\hfil \nobreak\hb@xt@\@pnumwidth{\hss #2}\par
\endgroup\fi}
\newcounter{sapp}[app]

\def\sapp{\def\theequation{\Alph{app}.\arabic{equation}}
\par
\@afterindentfalse
\secdef\@sapp\@dsapp}
\newcommand{\@sapp}{\@startsection{sapp}{2}{\z@}%
{-3.25ex\@plus -1ex \@minus -.2ex}%
{1.5ex \@plus .2ex}%
{\normalfont\large\bfseries}}

\def\@dsapp#1{%
{\parindent \z@ \raggedright \bf #1}\par\nobreak}
\newcommand{\l@sapp}{\@dottedtocline{2}{1.5em}{2.3em}}

\def\d{\partial}

\def\f{1\over}

\def\2{{1\over 2}}
\def\N2{${\cal N}=2$}

\def\be{ \begin{eqnarray} }
\def\ee{ \end{eqnarray} }

\def\d{\partial}

\def\bea{\begin{eqnarray}}
\def\eea{\end{eqnarray}}

\def\beq{\begin{equation}}
\def\eeq{\end{equation}}
\def\ba{\beq\new\begin{array}{c}}
\def\ea{\end{array}\eeq}
\def\be{\ba}
\def\ee{\ea}

\def\f{1\over}

\def\R{Ruijsenaars-Schneider\ }
\def\SW{Seiberg-\-Witten theory\ }

\def\N{{cal N}}
\begin{document}
\setcounter{footnote}{0}
\begin{center}
\hfill FIAN/TD-27/00\\
\hfill ITEP/TH-59/00\\
\hfill hep-th/0010078\\
\vspace{0.3in}
{\Large\bf Seiberg-Witten Theories, Integrable Models}
\medskip
{\Large\bf and
Perturbative Prepotentials}
\end{center}
\centerline{{\large A.Mironov}\footnote{Theory Dept.,
Lebedev Physical Inst. and ITEP, Moscow,
Russia; e-mail: mironov@itep.ru, mironov@lpi.ru}}
\bigskip
\abstract{\footnotesize
This is a very brief review of relations between Seiberg-Witten
theories and integrable systems
with emphasis on the perturbative prepotentials presented at the
E.S.Fradkin Memorial Conference.
}
\begin{center}
\rule{5cm}{1pt}
\end{center}
\bigskip
\setcounter{footnote}{0}
\paragraph{Introductory remarks.}
\SW \cite{SW} provides a description of the effective low-energy actions of
four
dimensional ${\cal N}=2$ SUSY $SU(N)$
Yang-Mills theories in terms of finite-dimensional integrable systems
\cite{GKMMM}.
Such a description has been extended to other ${\cal N}=2$ gauge theories.
There are basically three
different ways to extend the original \SW.
First of all, one may consider other gauge groups, from other simple
classical groups to those being a product of several simple factors.
The other possibility is to add some matter hypermultiplets in different
representations. The two main cases here are the matter in fundamental or
adjoint representations. At last, the third possible direction
to deform \SW is to consider 5- or 6-dimensional theories, compactified
respectively onto the circle of radius $R_5$ or torus with modulus $R_5/R_6$
(if the number of dimensions exceeds 6, the gravity becomes
obligatory coupled to the gauge theory).
It is important to investigate thoroughfully different deformations of \SW
and
corresponding deformations of the proper integrable systems in order to
better establish the correspondence between these two. In fact, since
repeating the
whole original Seiberg-Witten procedure, which might unquestionably prove
the correspondence, is technically quite tedious for
theories with more vacuum moduli, in most cases the procedure of
identification of the gauge theory and integrable model comes via comparing
the three characteristics:
\begin{itemize}
\item deformations of the two
\item number of the vacuum moduli and external parameters
\item perturbative prepotentials
\end{itemize}
The number of vacuum moduli (i.e. the number of scalar fields that may
have non-zero
v.e.v.`s) on the physical side should be compared with dimension of
the moduli space
of the spectral curves in integrable systems, while the external parameters
in
gauge theories (hypermultiplet masses) should be also some external
parameters
in integrable models (coupling constants, values of Casimir functions for
spin chains etc).
As for third item, the low energy effective action in \SW is described
in terms of {\it prepotentials} that celebrate a lot of
properties familiar from the original studies of pure topological theories
(where have been neglected the possibility of the light excitations to
move).
These properties include the identification \cite{GKMMM,IM2,RG} of
prepotentials as quasiclassical (Whitham)
$\tau$-functions \cite{Kri,Dub} and peculiar equations, of which the
(generalized) WDVV equations \cite{WDVV,Dub,MMM} are the best known example.
In the prepotential, the contributions of particles and
solitons/mo\-no\-po\-les (dyons) sharing the same mass scale, are still
distinguishable, because of different dependencies on the bare coupling
constant, {\it i.e.} on the modulus $\tau$ of the bare coordinate elliptic
curve (in the UV-finite case) or on the $\Lambda_{QCD}$ parameter (emerging
after dimensional transmutation in UV-infinite cases).  In the limit $\tau
\rightarrow i\infty$ ($\Lambda_{QCD} \rightarrow 0$), the solitons/monopoles
do not contribute and the prepotential reduces to the ``perturbative'' one,
describing the spectrum of non-interacting {\it particles}.  It is
immediately given by the SUSY Coleman-Weinberg formula \cite{WDVVlong}:
\be\label{ppg}
{\cal F}_{pert}(a) =
\sum_{\hbox{\footnotesize reps}
\ R,i} (-)^F {\rm Tr}_R (a + M_i)^2\log (a + M_i)
\ee
Seiberg-Witten theory (actually, the identification of appropriate
integrable
system) can be used to construct the non-perturbative prepotential,
describing the mass spectrum of all the ``light'' (non-stringy) excitations
(including solitons/monopoles).
Switching on Whitham times \cite{RG} presumably allows one to extract
some correlation functions in the ``light'' sector.
Now note that the problem of calculation of the prepotential in physical
theory is simple only at the perturbative level, where it is given just by
the leading contribution,
since the $\beta$-function in ${\cal N}=2$ theories is non-trivial only in
one loop.
However, the calculation of higher (instantonic) corrections in the gauge
theory can be hardly
done at the moment\footnote{In order to check the very ideology that
the integrable systems lead to the correct answers, there were
calculated first several corrections \cite{Instanton}. The results proved to
exactly
coincide with the predictions obtained within the integrable approach.}.
Therefore, the standard way of doing is to rely on integrable calculations.
This is why the establishing the ``gauge theories $\leftrightarrow$
integrable theories" correspondence is of clear practical
(apart from theoretical) importance.
\paragraph{Integrability of \SW.}
Let us describe now the correspondence in more details. The most
important result of
\cite{SW}, from this point of view, is that the moduli space of
vacua and low energy effective action in SYM theories
are completely given by the following input data:
\begin{itemize}
\item
Riemann surface ${\cal C}$
\item
moduli space ${\cal M}$ (of the curves ${\cal C}$)
\item
meromorphic 1-form $dS$ on ${\cal C}$
\end{itemize}
How it was pointed out in \cite{GKMMM,WDVVlong},
this input can be naturally described
in the framework of some underlying integrable system.
To this end, first, we introduce bare spectral curve $E$ that is torus
$y^2=x^3+g_2x^2+g_3$ for the UV-finite\footnote{The situation is still
unclear in application to the case of fundamental matter with the number
of matter hypermultiplets equal to $N_f =
2N$. In existing formulation for spin chains the bare coupling constant
appears rather as a twist in gluing the ends of the chain together
\cite{GGM1} (this parameter occurs only when $N_f = 2N$) and is not
immediately identified as a modulus of a {\it bare} elliptic curve. This
problem is a fragment of a more general puzzle:  spin chains have not been
described as Hitchin systems; only the ``$2\times 2$'' Lax representation is
known, while its ``dual'' $N_c\times N_c$ one is not yet available.}
gauge theories with the associated holomorphic 1-form
$d\omega=dx/y$. This bare spectral curve degenerates into the
double-punctured sphere (annulus) for the asymptotically free theories
(where dimensional transmutation occurs): $x\to
w+1/w$, $y\to w-1/w$, $d\omega=dw/w$.
On this bare curve, there are given either a
matrix-valued Lax operator $L(x,y)$ if one considers an extension of the
$N\times N$ Toda Lax representation, or another matrix Lax operator ${\cal
L}_i(x,y)$ associated with an extension of the $2\times 2$ Toda
Lax representation
and defining the transfer matrix $T(x,y)$. The corresponding dressed
spectral
curve ${\cal C}$ is defined either from the formula $\det(L-\lambda)=0$, or
from
$\det(T-w)=0$.
This spectral curve is a
ramified covering of $E$ given by the equation
\be
{\cal P}(\lambda;x,y)=0
\ee
In the case of the gauge group  $G=SU(n)$, the function ${\cal P}$ is a
polynomial of degree $n$ in $\lambda$.
Thus, we have the spectral curve ${\cal C}$ the moduli space ${\cal M}$ of
the
spectral curve given just  by
coefficients of ${\cal P}$.
The third important ingredient of the construction is the
generating 1-form $dS \cong \lambda d\omega$ meromorphic on
${\cal C}$ (``$\cong$" denotes the equality modulo total derivatives).
From the point of view of
the integrable system, it is just the shortened action "$pdq$" along the
non-contractible contours on the Hamiltonian tori. Its defining property is
that the derivatives of $dS$ with respect to the moduli (ramification
points)
are holomorphic differentials on the spectral curve.
The prepotential ${\cal F}$
and other ``physical" quantities are defined in terms of the
cohomology class of $dS$:
\be
a_i = \oint_{A_i} dS,\\ a_i^D\equiv {\d {\cal F}\over\d
a_i}=\oint_{B_i}dS,\\
\ A_I \circ B_J = \delta_{IJ}.
\label{defprep}
\ee
The first identity defines here the appropriate flat moduli, while the
second
one -- the prepotential. The defining property
of the generating differential
$dS$ is that its derivatives w.r.t. moduli
give holomorphic 1-differentials. In particular,
\be
{\d dS\over \d a_i}=d\omega_i
\ee
and, therefore, the second derivative of the prepotential
w.r.t. $a_i$'s is the period
matrix of the curve ${\cal C}$ (physically charges, i.e. effective
coupling constants in the gauge theory):
\be
{\d^2{\cal F}\over\d a_i\d a_j}=T_{ij}
\ee
The latter formula allows one to identify prepotential with logarithm of the
$\tau$-function of the Whitham hierarchy \cite{Kri,Dub,RG}: ${\cal
F}=\log\tau$.
So far we reckoned without matter hypermultiplets.
In order to include them, one just needs to consider the
surface ${\cal C}$ with punctures. Then, the masses are proportional to the
residues of $dS$ at the punctures, and the moduli space has to be extended
to
include these mass moduli. All other formulas remain in essence the same
(see \cite{WDVVlong} for more details).
\paragraph{Seiberg-Witten theories vs. integrable systems.}
The table of known relations between gauge theories and integrable systems
is
drawn in Fig.2.
\begin{figure}[t]
\begin{center} \begin{tabular}{|c|c|c|c|c|}
\hline
{\bf SUSY}&{\bf Pure gauge}&{\bf SYM theory}&{\bf SYM theory}\\
{\bf physical}&{\bf SYM theory,}&{\bf with fund.}&{\bf with adj.}\\
{\bf theory}&{\bf gauge group $G$}&{\bf matter}&{\bf matter}\\
\hline
        & inhomogeneous  & elliptic  & inhomogeneous        \\
         & periodic        & Calogero & periodic \\
{\bf 4d} & Toda chain & model & $XXX$ \\
& for the dual affine ${\hat G}^{\vee}$  &({\it trigonometric}& spin chain\\
        & ({\it non-periodic}  & {\it Calogero}& ({\it non-periodic}\\
        & {\it Toda chain})& {\it model})& {\it chain})\\
\hline
        & periodic  & elliptic & periodic   \\
        & relativistic  & Ruijsenaars&$XXZ$\\
{\bf 5d} & Toda chain  & model & spin chain  \\
        & ({\it non-periodic} & ({\it trigonometric}& ({\it non-periodic}\\
        & {\it chain}) & {\it Ruijsenaars})&{\it chain})\\
\hline
    &periodic & Dell & periodic\\
    & ``Elliptic" & system & $XYZ$ (elliptic)\\
{\bf 6d} &  Toda chain & ({\it dual to elliptic} & spin chain\\
        & ({\it non-periodic}& {\it Ruijsenaars,}&({\it non-periodic} \\
& {\it chain})& {\it elliptic-trig.})&{\it chain})\\
\hline
\end{tabular}
\end{center}
\caption{SUSY gauge theories $\Longleftrightarrow$ integrable systems
correspondence. The perturbative limit is marked by the italic font (in
parenthesis).}\label{intvsYM}
\end{figure}
\vspace{10pt}
The original Seiberg-Witten model, which is the $4d$ pure gauge $SU(N)$
theory
(in fact, in their papers \cite{SW}, the authors considered the $SU(2)$
case only, but the generalization made in \cite{Theisen} is quite
immediate), is the upper left square of the table. The
remaining part of the table contains possible deformations.
Here only two of the three possible ways to deform the original
Seiberg-Witten
model are shown. Otherwise, the table would be three-dimensional.
In fact,the third direction related to changing the gauge group,
although being of an interest is slightly out of the main line. Therefore,
we only make several comments on it. In
particular, right here note that the generalization of the original
\SW to another simple group $G$ is quite immediate and corresponds to
the Toda chain for the dual affine algebra
${\hat G}^{\vee}$ \cite{MW}.
One direction in the table corresponds to matter hypermultiplets
added. The most interesting is to add matter in adjoint or fundamental
representations, although attempts to add antisymmetric and
symmetric matter hypermultiplets were also done (see \cite{asym} for
the construction of the curves and \cite{KPasym} for the corresponding
integrable systems). Adding matter in other representations in the basic
$SU(N)$ case leads to non-asymptotically free theories.
\underline{Columns: Matter in adjoint {\it vs.} fundamental
representations of the gauge group.}
Matter in adjoint representation can be described in terms of a
larger pure SYM model, either with higher SUSY or in higher space-time
dimension.  Thus models with adjoint matter form a hierarchy, naturally
associated with the hierarchy of integrable models {\it Toda chain
$\hookrightarrow$ Calogero $\hookrightarrow$
Ruijsenaars $\hookrightarrow$ Dell}
\cite{GKMMM,intadj,IM1,IM2,BMMM2,BMMM3,MM1,MM2}\footnote{The
generalization of the
models of this class to other groups can be found in \cite{DPE}.
}. Similarly, the
models with fundamental matter are related to the hierarchy of spin chains
originated from the Toda chain: {\it Toda chain
$\hookrightarrow$ XXX $\hookrightarrow$
XXZ $\hookrightarrow$ XYZ}
\cite{XXX,XYZ,GGM2,MarMir}\footnote{The other classical groups for
spin chains/fundamental matter theories were considered in \cite{GM}.
Another possibility, the gauge groups which are products of simple factors
and bi-fundamental matter, were proposed in \cite{W}, while the
corresponding integrable models discussed in \cite{GGM1}.}.
Note that, while coordinates in integrable systems describing pure gauge
theories and those with fundamental matter, live on the cylinder (i.e.
the dependence on coordinates is trigonometric), the coordinates in the
Calogero system (adjoint matter added)
live on a torus\footnote{Since these theories are
UV finite, they depend on an additional (UV-regularizing) parameter, which
is exactly the modulus $\tau$ of the torus.}. However, when one takes the
perturbative limit, the coordinate dependence becomes trigonometric.
\underline{Lines: Gauge theories in different dimensions.}
Integrable systems relevant
for the description of vacua of $d=4$ and $d=5$ models are
respectively the Calogero and Ruijsenaars ones (which possess the ordinary
Toda chain and ``relativistic Toda chain'' as Inosemtsev's limits
\cite{Ino}), while $d=6$ theories are described by the
double elliptic (Dell) systems. When we go from $4d$ (Toda, $XXX$, Calogero)
theories to $5d$ (relativistic Toda, $XXZ$, Ruijsenaars) theories
the momentum-dependence of the Hamiltonians
becomes trigonometric instead of rational. Similarly, the $6d$ theories
give rise to an elliptic momentum-dependence of the Hamiltonians. Since
adding
the adjoint hypermultiplet elliptizes the coordinate dependence, the
integrable
system corresponding to $6d$ theory with adjoint matter celebrates both
the coordinate- and momentum-dependencies elliptic. A candidate for ``the
ellitpic Toda chain'' was proposed in \cite{Kriel}.
\paragraph{Prepotentials in Seiberg-Witten and integrable theories.}
We already discussed the important role of the prepotentials, in
particular, of their perturbative part. This latter helps one to make
an identification of the integrable theory with the gauge theory.
Note that many crucial properties of the prepotential
are seen already at the perturbative
level (say, the WDVV equations \cite{WDVVlong,Mir99}). Therefore, now we
are going to give an explicit description of the perturbative prepotentials
for the Seiberg-Witten theories listed in the previous paragraph.
The technical tool that allows one to proceed with effective
perturbative expansion of the prepotential is ``the residue formula'',
the variation of the period matrix, that is the third derivatives
of ${\cal F}(a)$ (see, e.g. \cite{WDVVlong}):
\be
\frac{\partial^3{\cal F}}{\partial a_i\partial a_j\partial a_k}
= \frac{\partial T_{ij}}{\partial a_k} ={1\over 2\pi i}
{\rm res}_{d\xi = 0} \frac{d\omega_id\omega_jd\omega_k}{\delta dS},
\label{res}
\ee
where $\delta dS\equiv d\left({dS\over d\xi}\right)d\xi $.
We remark that although
$d\xi$ does not have zeroes on the {\it bare} spectral
curve when it is a torus or doubly punctured sphere, it does
in general however possess them on the covering ${\cal C}$.
In order to construct the perturbative prepotentials, one merely
can note that, at the leading order, the Riemann surface (spectral
curve of the integrable system) becomes rational. Therefore, the
residue formula allows one to obtain immediately the third derivatives
of the prepotential as simple residues on the sphere. In this way,
one can check
that the prepotential has actually the form (\ref{ppg}) (see details in
\cite{WDVVlong}).
As a concrete example, let us consider the $SU(n)$ gauge group. Then, say,
perturbative prepotential for the pure gauge theory acquires the
form
\be
{\cal F}_{pert,V}={\f 4}\sum_{ij}
\left(a_i-a_j\right)^2\log\left(a_i-a_j\right)
\ee
This formula establishes that when v.e.v.'s
of the scalar fields in the gauge supermultiplet are non-vanishing
(perturbatively $a_r$ are eigenvalues of the vacuum
expectation matrix  $\langle\phi\rangle$), the fields in the gauge multiplet
acquire masses $m_{rr'} = a_r - a_{r'}$ (the pair of indices $(r,r')$ label
a field in the adjoint representation of $G$). In the $SU(n)$ case,
the eigenvalues are subject to the condition $\sum_ia_i=0$.
Analogous formula for the
adjoint matter contribution to the prepotential is
\be
{\cal F}_{pert,A}=-{\f 4}\sum_{ij}
\left(a_i-a_j+M\right)^2\log\left(a_i-a_j+M\right)
\ee
while the contribution of one fundamental matter hypermultiplet reads as
\be
{\cal F}_{pert,F}=-{\f 4}\sum_{i}
\left(a_i+m\right)^2\log\left(a_i+m\right)
\ee
Similar formulas can be obtained for the other groups.\footnote{The
eigenvalues of $\langle\phi\rangle$
in the first
fundamental representation of the classical series of the Lie groups are
\be
B_n\ (SO(2n+1)):\ \ \ \ \ \{a_1,...,a_n,0,-a_1,...,-a_n\};\\
C_n\ (Sp(n)):\ \ \ \ \ \{a_1,...,a_n,-a_1,...,-a_n\};\\
D_n\ (SO(2n)):\ \ \ \ \ \{a_1,...,a_n,-a_1,...,-a_n\}
\ee
while the eigenvalues in the adjoint representation have the form
\be\label{adj}
B_n:\ \ \ \ \ \{\pm a_j;\pm a_j\pm a_k\};\ \ \ j<k\le n\\
C_n:\ \ \ \ \ \{\pm 2a_j;\pm a_j\pm a_k\};\ \ \ j<k\le n\\
D_n:\ \ \ \ \ \{\pm a_j\pm a_k\}, \ \ j<k\le n
\ee
Analogous formulas can be written for the exceptional groups too.  The
prepotential in the pure gauge theory can be
read off from the formula (\ref{adj}) and has the form
\be\label{adjvo}
B_n:\ \ \ \ \ {\cal F}_{pert}={1\over 4}\sum_{i,i} \left(\left(a_{i}-a_{j}
\right)^2\log\left(a_{i}-a_{j}\right)+\left(a_{i}+a_{j}
\right)^2\log\left(a_{i}+a_{j}\right)\right)+
{1\over 2}\sum_i a_i^2\log a_i;\\
C_n:\ \ \ \ \ {\cal F}_{pert}={1\over 4}\sum_{i,i} \left(\left(a_{i}-a_{j}
\right)^2\log\left(a_{i}-a_{j}\right)+\left(a_{i}+a_{j}
\right)^2\log\left(a_{i}+a_{j}\right)\right)
+2\sum_i a_i^2\log a_i;\\
D_n:\ \ \ \ \ {\cal F}_{pert}={1\over 4}\sum_{i,i} \left(\left(a_{i}-a_{j}
\right)^2\log\left(a_{i}-a_{j}\right)+\left(a_{i}+a_{j}
\right)^2\log\left(a_{i}+a_{j}\right)\right)
\ee
} The perturbative prepotentials in $4d$ theories with fundamental matter
are discussed in detail in \cite{WDVVlong} (see also \cite{KDP1}).
The formulas presented above in the $4d$ case
can be almost immediately extended to the
``relativistic"\footnote{i.e. with trigonometric momentum dependence}
 $5d$ \N2 SUSY gauge models
with one compactified dimension. One can understand the
reason for this ``relativization'' in the following way. Considering
four plus one compact dimensional theory one should take into account the
contribution of all Kaluza-Klein modes to each 4-dimensional field.
Roughly speaking it leads to the 1-loop contributions to the effective
charge of the form\footnote{Hereafter, we denote $a_{ij}\equiv a_i-a_j$.}
\be\label{relKK}
T_{ij} \sim \sum _{\rm masses}\log\hbox{ masses} \sim \sum _m\log
\left(a_{ij} +
{m\over R_5}\right) \sim \log\prod _m\left(R_5a_{ij} + m\right)
\sim\log\sinh R_5a_{ij}
\ee
i.e. coming from $4d$ to $5d$
one should make a substitution $a_{ij} \rightarrow
\sinh R_5a_{ij}$, at least, in the
formula for the perturbative prepotential.
Now the same general argument applied to the $5d$ case,
can be equally applied to the $6d$ case,
or to the theory with {\em two} extra compactified dimensions,
of radii $R_5$ and $R_6$. Indeed, the account of
the Kaluza-Klein modes allows one to predict the perturbative form
of charges in the $6d$ case as well. Namely, one should expect them
to have the form\footnote{Hereafter, $\theta_*$ denotes the
$\theta$-function
with odd characteristics.}
\be\label{6dKK}
T_{ij} \sim \sum _{\rm masses}\log\hbox{ masses} \sim \sum _{m,n}\log
\left(a_{ij} +
{m\over R_5} +{n\over R_6}
\right) \sim
\\
\sim\log\prod _{m,n}\left(R_5a_{ij} + m+n{R_5\over R_6}\right)
\sim\log\theta_*\left(R_5a_{ij}\left|i{R_5\over R_6}\right)\right.
\ee
i.e. coming from $4d$ ($5d$) to $6d$
one should replace the rational (trigonometric) expressions by the elliptic
ones, at least, in the
formulas for the perturbative prepotential, the (imaginary part of) modular
parameter being
identified with the ratio of the compactification radii $R_5/R_6$.
More rigid derivations
using the residue formula confirms these results. In the fundamental matter
case, they look as follows \cite{WDVVlong,GGM2,MarMir}.
The spectral curve in all cases can be written in the form
\be\label{scg}
w+{Q^{(d)}(\xi)\over w}=2P^{(d)}(\xi)
\ee
or
\be\label{scg'}
W+{1\over W}={2P^{(d)}(\xi)\over\sqrt{Q^{(d)}(\xi)}},
\ \ \ \ \ \ W \equiv {w\over\sqrt{Q^{(d)}(\xi)}}
\ee
In the perturbative limit, only the first term in the r.h.s.
of these formulas survives.
The generating differential $dS$ is always of the form
\be\label{dSr}
dS=\xi d\log W
\ee
The concrete forms of the functions introduced here are:
\be
Q^{(4)}(\xi)\sim\prod_\alpha^{N_f}(\xi-m_\alpha),\ \ \
Q^{(5)}(\xi)\sim\prod_\alpha^{N_f}\sinh(\xi-m_\alpha),\ \ \
Q^{(6)}(\xi)\sim\prod_\alpha^{N_f}{\theta_*(\xi-m_\alpha)\over\theta_*^2
(\xi-\xi_i)}
\ee
\be
P^{(4)}\sim\prod_i^{N}(\xi-a_i),\ \ \
P^{(5)}\sim\prod_i^{N}\sinh(\xi-a_i),\ \ \
P^{(6)}\sim\prod_i^{N}{\theta_*(\xi-a_i)\over\theta_*(\xi-\xi_i)}
\ee
(in $P^{(5)}(\xi)$, there is also some exponential of $\xi$ unless $N_f=2N$,
see \cite{MarMir}), $\xi_i$`s being just external parameters.
The perturbative part of the prepotential can be calculated using
these manifest expressions and the residue formula and
is always of the
form\footnote{The term in the prepotential that depends only on
masses is not essential for the standard \SW but is crucial for
the prepotential to enjoy its main properties, similar to the WDVV
equations. This term is unambigously restored from the residue formula.}
\be\label{gp}
{\cal F}_{pert}={1\over 4}\sum_{i,j}f^{(d)}(a_{ij})-{1\over
4}\sum_{i,\alpha}
f^{(d)}(a_i-m_{\alpha})+{1\over 16}\sum_{\alpha,\beta}
f^{(d)}(m_{\alpha}-m_{\beta})
\ee
The explicit form of these functions is
\be
f^{(4)}(x)=x^2\log x,\ \ \
f^{(5)}(x)=\sum_{n}f^{(4)}\left(x+{n\over R_5}\right)=
{1\over 3}\left|x^3
\right|-{1\over 2}{\rm Li}_3\left(e^{-2|x|}\right),\\
f^{(6)}(x)=\sum_{m,n}f^{(4)}\left(x+\frac{n}{R_{5}}+\frac{m}{R_{6}}\right)=
\sum_n f^{(5)}\left(x+n{R_{5}\over R_6}\right)=
\\
= \left({1
\over 3}\left|x^3\right|
-{1\over 2}{\rm Li}_{3,q}\left(e^{-2|x|}\right)
+ {\rm quadratic\ \ terms}\right)
\ee
so that
\be
{f^{(4)}}''=\log x,\ \ \ {f^{(5)}}''(x)=\log\sinh x,\ \ \
{f^{(6)}}''(x)=\log\theta_*(x)
\ee
Note that, in the $6d$ case,
$N_f$ is always equal to $2N$,\footnote{Otherwise, the function in the
r.h.s. of formula (\ref{scg'}) is not elliptic.} and, in $d=5,6$,
there is a restriction $\sum a_i=\sum\xi_i=\2\sum m_\alpha$
which implies that the gauge moduli would be
rather associated with $a_i$ shifted by the constant
${1\over 2N}\sum m_\alpha$.
In these formulas, ${\rm Li}_{3}(x)$  is the tri-logarithm, while
${\rm Li}_{3,q}(x)$  is the elliptic tri-logarithm
\cite{mamont}.
Now let us consider the adjoint matter case\footnote{The prepotential in
the pure gauge theory
is described by the first term in (\ref{gp}) in $d=4$, while in
$d=5$ it gets an additional cubic term
\be\label{FARTC}
{\cal F}_{pert}={1\over 4}\sum_{i,j}\left({1
\over 3}a_{ij}^3
-{1\over 2}{\rm Li}_3\left(e^{-2a_{ij}}\right)\right)+
{N_c\over 2}\sum_{i>j>k}a_ia_ja_k=
\\={1\over 4}\sum_{i,j}\left({1
\over 3}a_{ij}^3
-{1\over 2}{\rm Li}_3\left(e^{-2a_{ij}}\right)\right)+
{N_c\over 6}\sum_{i}a_i^3
\ee
Because of the requirement $N_f=2N$, it is still unclear how to deal
with the pure gauge ($N_f=0$)
theory in $d=6$.}. In this case, using the previously developed general
arguments, one would expect for the perturbative prepotential to be of the
form
\be\label{app}
{\cal F}_{pert}={1\over 4}\sum_{i,j}f^{(d)}(a_{ij})-
{1\over 4}\sum_{i,j}f^{(d)}(a_{ij}+M)
\ee
However, the calculations in this case are more complicated.
In particular, the spectral curves are far more involved (see
\cite{IM1,KDP,BMMM2,BMMM3,MM2}). Nevertheless, in the perturbative limit
they
simplify drastically and coincide with the perturbative curves for
the fundamental matter case with the hypermultiplet masses
pairwise identified and equal to $a_i+M$, with the generating 1-form
$dS$  being of the same form (\ref{dSr}).
This immediately leads to the result (\ref{app})
\cite{KDP,BMMM1,BMMM2,MM1,MM2}.
I am grateful to H.W.Braden, A.Gorsky, A.Marshakov and A.Morozov for
numerous discussions and T.Takebe for kind hospitality at the
Ochanomizu University, Tokyo, where this work was completed. The work
is partly supported by the grants: RFBR-00-02-16477-a,
INTAS 99-0590, CRDF grant \#6531 and the JSPS fellowship for research in
Japan.

\end{document}